\documentclass[aps,prl,showpacs,amsmath,superscriptaddress,twocolumn,floatfix,nobalancelastpage]{revtex4-1}

\usepackage[nottoc,numbib]{tocbibind}

\usepackage[colorlinks=true, citecolor=black, linkcolor=black, urlcolor=black]{hyperref}

\usepackage[all]{hypcap}
\usepackage{amsmath}
\usepackage{enumerate,bm}
\usepackage{graphicx}
\usepackage{grffile} 
\usepackage{color}
\usepackage{esint}
\usepackage{textpos}
\usepackage[usenames,dvipsnames]{xcolor}
\usepackage{subfigure}
\usepackage{flushend}
\usepackage{tabularx}
\usepackage{amssymb}
\usepackage{float}
\usepackage{subfigure}
\usepackage{ulem} 

\usepackage{setspace}
\usepackage{verbatim}

\begin{document}

\author{Simone~Latini}
\email{simone.latini@mpsd.mpg.de}
\affiliation{Max Planck Institute for the Structure and Dynamics of Matter and Center for Free Electron Laser Science, 22761 Hamburg, Germany}

\author{Umberto~De~Giovannini}
\email{umberto.degiovannini@gmail.com}
\affiliation{Max Planck Institute for the Structure and Dynamics of Matter and Center for Free Electron Laser Science, 22761 Hamburg, Germany}
\affiliation{Nano-Bio Spectroscopy Group,  Departamento de Fisica de Materiales, Universidad del País Vasco UPV/EHU- 20018 San Sebastián, Spain}

\author{Edbert J. Sie}
\affiliation{Department of Physics, Massachusetts Institute of Technology, Cambridge, MA 02139, USA}
\affiliation{Geballe Laboratory for Advanced Materials, Stanford University, Stanford, CA 94305, USA}

\author{Nuh Gedik}
\affiliation{Department of Physics, Massachusetts Institute of Technology, Cambridge, MA 02139, USA}

\author{Hannes~H\"ubener}
\email{hannes.huebener@gmail.com}
\affiliation{Max Planck Institute for the Structure and Dynamics of Matter and Center for Free Electron Laser Science, 22761 Hamburg, Germany}

\author{Angel~Rubio}
\email{angel.rubio@mpsd.mpg.de}
\affiliation{Max Planck Institute for the Structure and Dynamics of Matter and Center for Free Electron Laser Science, 22761 Hamburg, Germany}
\affiliation{Nano-Bio Spectroscopy Group,  Departamento de Fisica de Materiales, Universidad del País Vasco UPV/EHU- 20018 San Sebastián, Spain}
\affiliation{Center for Computational Quantum Physics (CCQ), The Flatiron Institute, 162 Fifth avenue, New York NY 10010.}

\title{Phonoritons in a monolayer hBN optical cavity}

\date{\today}

\begin{abstract}
A phonoriton is an elementary excitation that is predicted to emerge from hybridization between exciton, phonon and photon. Besides the intriguing many-particle structure, phonoritons are of interest as they could serve as functional nodes in devices that utilize electronic, phononic and photonic elements for energy conversion and thermal transport applications. Although phonoritons are predicted to emerge in an excitonic medium under intense electromagnetic wave irradiation, the stringent condition for their existence has eluded direct observation in solids. In particular, on-resonance, intense pumping scheme has been proposed but excessive photoexcitation of carriers prevents optical detection. Here we theoretically predict the appearance of phonoritonic features in monolayer hexagonal boron nitride (hBN) embedded in an optical cavity. The coherent superposition nature of phonoriton states is evidenced by the hybridization of exciton-polariton branches with phonon replicas that is tunable by the cavity-matter coupling strength. This finding simultaneously provides an experimental pathway to observe the predicted phonoritons and opens a new avenue for tuning materials properties.
\end{abstract}
\maketitle

New states of matter can arise from hybridization of fundamental modes of solid state systems with light. When the light-matter interaction is strong, for instance through confinement of the photons in optical resonators or cavities, these states can be described as bosonic quasiparticles, polaritons.  There has been a resurgence of interest to the change of the properties in the material that host the polariton inside the cavity. Especially changes to the electron-phonon coupling and hence the possibility to affect superconducting properties of materials in optical cavities with strong coupling are being explored theoretically, predicting an enhancement of the transition temperature~\cite{Curtis:2019kj}, control of electron-phonon parameter~\cite{Sentef:2018gp} or even an altogether new electron pairing mechanism through cavity photons~\cite{Schlawin:2019jw}, while first experimental evidence points towards the possibility to enhance the superconducting gap through cavity coupling~\cite{AnoopARXIVE2019}. Among other features that have been observed for materials in strongly coupled cavities are enhanced exciton lifetimes~\cite{ChenNatPhoton2017,SunNatPhoton2017,DufferwielNatPhoton2017}, theoretical predictions for controlling of exciton ordering~\cite{Latini:2019bz}, an exciton-insulator superradiance phase~\cite{Mazza2019} as well as the possibility to control the para- to ferroelectric phase transition in SrTiO$_3$~\cite{Ashida:2020,Latini:2021}. Finally, the possibility to break time-reversal symmetry with circular polarized cavity photons has been suggested to allow control of topological properties~\cite{Hubener:2020fm}. These recent developments suggest the emergence of the field of cavity materials engineering. 

When changing materials properties with cavities, one can in principle couple any elementary excitation of the solid to the cavity photons, in accordance with the known "zoo" of solid state polaritons~\cite{Basov:2016eq}, including exciton-polaritons and phonon-polaritons. Even though such excitations occur in real materials at very different energy scales, through the coupling to a cavity, they can be tuned into mutual resonance. An example of a three-component polarization quasiparticle as the hybridization between exciton, phonon and photon has been predicted by Ivanov and Keldysh~\cite{KeldyshJETPLett1979} to emerge in a semiconductor under intense electromagnetic radiation tuned at the exciton resonance. The authors called this new quasiparticle a \textit{phonoriton} to highlight its status as a fundamental mode on par with polaritons. However, despite the theoretical proposal, there has been little experimental evidence, except for a few reports~\cite{Vygovskii:1985, Greene:1988, Hanke:1999} that implicitly invoked this phenomena, because direct observation in laser pumped systems is experimentally challenging. In particular, the on-resonance, intense pumping scheme leads to excessive photoexcitation of carriers that prevents the optical detection, and may possibly damage the materials before the phonoritons start to emerge. By contrast, in a cavity the strong light-matter interaction is not achieved via intensity of the light, but instead by an enhanced coupling strength that results from confinement of the photon modes.

\begin{figure}
  \centering
   \includegraphics[width=\columnwidth]{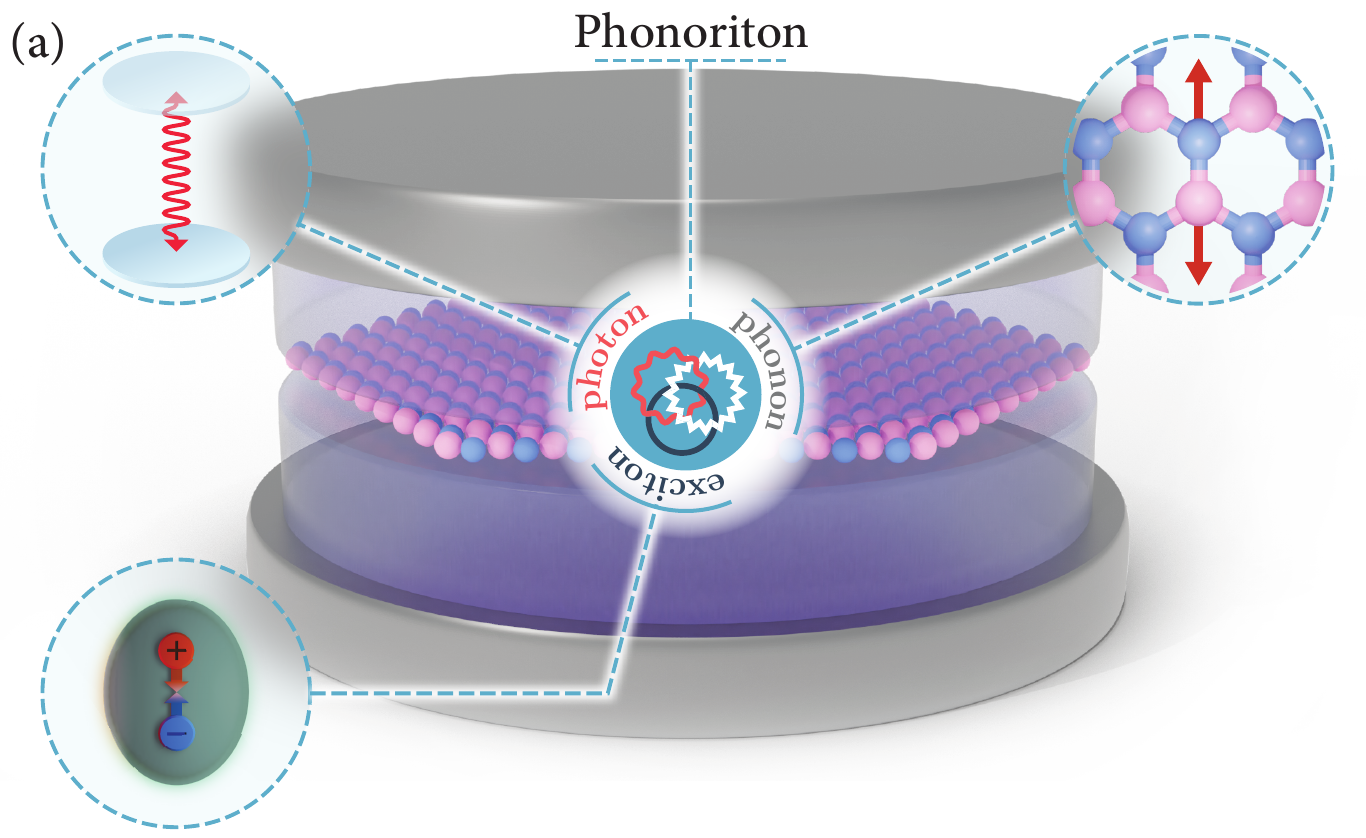}
   \caption{\label{fig:1}
   The phonoriton is a hybridization between excitons, phonons and photons that can be realized by embedding a crystal in an optical cavity. }
\end{figure}

Here we predict from first principles the appearance of such a three-component composite quasiparticle in a monolayer hBN embedded in an optical cavity and  discuss its realization in other 2D materials. By combining many-body perturbation theory and the quantum electrodynamical description of the exciton-polariton~\cite{Latini:2019bz} we show that phonoriton states occur when Rabi splitting of the polariton comes into resonance with phonon modes of the material. Such states are neither purely excitonic nor phononic but the presence of the cavity photons leads to a mixing of all three modes as sketched in Fig.~\ref{fig:1}. We argue that the realization of this proposal gives in fact a novel view on the realities of materials in optical cavities and the prospective control and design of their properties. The tunability of elementary modes by dressing with light allows to create new particle hybridizations and provides an opportunity to explore new directions of materials research utilizing strong light-matter interaction as well as providing a fruitful platform to rethink fundamental excitations in solids. 

To describe  the dressing of excitons by cavity photons from first principles, we consider the fundamental many-body Hamiltonian of mutually interacting electrons, and photons in (non-relativistic) quantum electrodynamics (QED). For a single photonic mode of energy $\omega_c$ in the velocity gauge representation and within the dipole approximation, reads:
\begin{equation}\label{eq:QEDgen}
\begin{split}
\hat{H}_{\rm QED} = &\hat{H}_{\rm{el}} + \omega_c \hat{b}^\dagger \hat{b} +  \frac{A_0^2}{2}(\hat{b}^\dagger + \hat{b})^2+\\
&+ A_0 \sum_{rs\bf{k}}\left(\langle\phi_{r\bf{k}} | \hat{e}\cdot\hat{p} | \phi_{s\bf{k}}\rangle \hat{c}^\dagger_{r\bf{k}}\hat{c}_{s \bf{k}} \hat{b}^\dagger +  h.c.\right),
\end{split}
\end{equation}
where $\hat{b}^\dagger$ and $\hat{b}$ are the photon creation and annihilation operators, $\hat{H}_{\rm{el}}$ the many-body electronic Hamiltonian, $\hat{c}^\dagger_{r\bf{k}},\hat{c}_{s \bf{k}}$ the electronic creation and annihilation operators (with $r,s$ band indices and $\bf{k}$ wavevectors in the first Brillouin zone), $\hat{p}$ the single particle momentum operator, $\phi$ the single electron wavefunctions, $\hat{e}$ the photon field polarization, and $A_0$ the amplitude of the vector potential. The generalization to a multi-mode setting is straightforward and can be found in Ref.~\cite{Latini:2019bz}.

To describe the effect of lattice motion and introduce the interaction with phonons we apply linear response theory and expand both the electronic and the light-matter term in the QED Hamiltonian in terms of the phononic displacement coordinates. By considering phonons $\alpha$ of energy $\Omega_{\alpha}$ and retaining the terms up to first order in the phonon displacement, the standard exciton-phonon coupling emerges as well as a three-way quasiparticle coupling arises from the exciton-photon term, a full derivation is provided in the SI. Since the coupling of photons to the electronic degrees of freedom involves the creation/annihilation of neutral electron-hole pairs, it is natural to approximate the many-body eigenstates of the electronic Hamiltonian by exciton states, as done in Ref.~\cite{Latini:2019bz}, i.e. $\hat{H}_{\rm{el}}|\Psi_{i}^{\rm{exc}}\rangle\simeq\epsilon_i^{\rm{exc}}|\Psi_{i}^{\rm{exc}}\rangle$. The phonon perturbed Hamiltonian, including the free phonon energy, reads in the excitonic basis:
\begin{equation}\label{eq:QED_corrected}
\begin{split}
\langle\Psi&_{i}^{\rm{exc}}|\hat{H}|\Psi_{j}^{\rm{exc}}\rangle = \\
&\left[\epsilon_{i}^{\rm{exc}} +\omega_c \hat{b}^\dagger \hat{b} + \sum_{\alpha}\Omega_{\alpha} \hat{a}^\dagger_{\alpha} \hat{a}_{\alpha} + \frac{A_0^2}{2}(\hat{b}^\dagger + \hat{b})^2\right]\delta_{ij}\\  &+\sum_{\alpha}\left(\mathcal{G}^{\rm exc}_{ij, \alpha} \hat{a}^\dagger_{\alpha} + \mathcal{G}^{\rm exc *}_{ji, \alpha} \hat{a}_{\alpha}\right) + A_0 \left(\mathcal{M}^{\rm exc}_{ij} \hat{b}^\dagger + \mathcal{M}^{\rm exc *}_{ji} \hat{b}\right)\\ 
&+ A_0 \sum_{\alpha}\left(\mathcal{Z}^{\rm exc}_{ij, \alpha} \hat{b}^\dagger + \mathcal{Z}^{\rm exc *}_{ji, \alpha} \hat{b}\right)\left(\hat{a}^\dagger_{\alpha} + \hat{a}_{\alpha})\right.
\end{split}
\end{equation}
where $\hat{a}^\dagger_{\alpha}$ and $\hat{a}_{\alpha}$ are the phonon (photon) creation and annihilation operators respectively, $\mathcal{M}^{\rm exc}_{ij} = \langle\Psi_{i}^{\rm{exc}}|\sum_{rs\bf{k}} \langle\phi_{r\bf{k}} | \hat{e}\cdot\hat{p} | \phi_{s\bf{k}}\rangle \hat{c}^\dagger_{r\bf{k}}\hat{c}_{s \bf{k}}|\Psi_{j}^{\rm{exc}}\rangle $
are the exciton-photon matrix elements, $\mathcal{G}^{\rm exc}_{ij, \alpha} =  \langle\Psi_{i}^{\rm{exc}}|\sum_{rs\bf{k}} g_{rs\bf{k}}^{\alpha}\hat{c}^\dagger_{r\bf{k}}\hat{c}_{s \bf{k}}|\Psi_{j}^{\rm{exc}}\rangle $
are the exciton-phonon matrix elements with $g$ the standard electron-phonon matrix elements, and $\mathcal{Z}_{ij}^{\rm exc}=\sqrt{\frac{1}{2M_{\alpha}\Omega_{\alpha}}}\langle\Psi_{i}^{\rm{exc}}|\sum_{rs\bf{k}} \frac{\partial}{\partial{\bf R}_{\alpha}}\langle\phi_{r\bf{k}} | \hat{e}\cdot\hat{p} | \phi_{s\bf{k}}\rangle\hat{c}^\dagger_{r\bf{k}}\hat{c}_{s \bf{k}}|\Psi_{j}^{\rm{exc}}\rangle$, with ${\bf R}_{\alpha}$ phonon displacement, describe the coupling between phonons, photons and excitons and we therefore refer to $\mathcal{Z}$ as the phonoriton matrix element. The Hamiltonian above is the central quantity of this letter and its eigenstates are the phonoriton quasiparticle states, $\Psi^{\rm p}_I$ discussed below, with $I$ the phonoriton index. The formation of a phonoriton can be understood, in simple terms, as the photon dressing of an exciton-phonon system which results in an new interaction term that couples the exciton and the phonons beyond the standard exciton-phonon coupling (see Fig.~S1 in the SI).
We stress that differently from the model of the original proposal ~\cite{KeldyshJETPLett1979, Wang:1990im}, here we do not involve scattering of higher momentum excitons and phonons. We explicitly account for the photon-phonon interaction only through the exciton since the relevant modes that we consider for the following phonoritonic features are not optically active. 

The Hamiltonian in Eq.~\ref{eq:QED_corrected} can be numerically diagonalized in a mixed exciton-photon-phonon product-state basis $|\Psi_{i}^{\rm{exc}}\rangle\otimes|n\rangle\otimes|m\rangle$, where $|n\rangle$ and $|m\rangle$ are phonon and photon number states respectively. We set up the Hamiltonian for the specific case of monolayer hBN, whose phonons have been extensively investigated in the case of phonon-polaritons formation \cite{dai2014tunable, low2017polaritons,serrano:2007}. To give a clear account of the mechanism driving the creation of the phonoriton, here we simplify the calculations by including the lowest excitonic state and by considering only the zero-momentum optical phonon modes. The in-plane modes in hBN are usually denoted as longitudinal (LO) and transverse (TO) modes indicating their contribution to a macroscopic polarization field, however at the Brillouion zone center they are degenerate at $\sim$166~meV. The out-of-plane mode (ZO) has an energy of $\sim$96~meV. Given the high energy of such phonons we neglect, in the following, the effect of thermal population.
The excitons and phonons are calculated from first-principles: for the exciton we solved the Bethe-Salpeter equation using the GPAW code~\cite{MortensenPhysRevB2005,EnkovaaraJPhysCondMat2010} while the electron-phonon matrix elements were computed with the Octopus code~\cite{TancogneDejean:2020ek}. From this we are able to calculate the matrix elements $\mathcal{M}^{\rm{exc}}_{mn}$, $\mathcal{G}^{\rm exc}_{ij, \alpha}$ and $\mathcal{Z}^{\rm{exc}}_{mn}$ in the Hamiltonian of Eq.~\ref{eq:QED_corrected} on a first principles level, for the specific case of monolayer hBN. For the phonoritonic matrix elements we assume that the envelope function of the exciton is, to first order, unchanged and we only account for the variation of the conduction-valence matrix elements as a function of lattice displacement. 
For the photons, we consider the first non-zero photon mode, $\omega_c=\pi c/L_{\perp}$, which has an out-of-plane wavevector and in-plane electric field and is therefore able to couple to the hBN in-plane excitonic dipole. Changing $L_{\perp}$, the vertical dimension of the cavity, allows for the tuning of the cavity frequency. This can be done by, for instance, adding an optically inactive spacer material. The quasi-2D cavity configuration is characterized by a frequency independent vector potential amplitude which for a perfect cavity with an embedding dielectric medium with dielectric constant $\epsilon=1$ is $A_0$, i.e. $A_0=1/\sqrt{a L_{\perp} \omega_c}=1/\sqrt{2\pi c~a}$, where $a$ is the unit cell area of hBN. In the following we investigate the effect of cavity-matter coupling by using $A_0$ as a tuning parameter.
\begin{figure}
  \centering
   \includegraphics[width=\columnwidth]{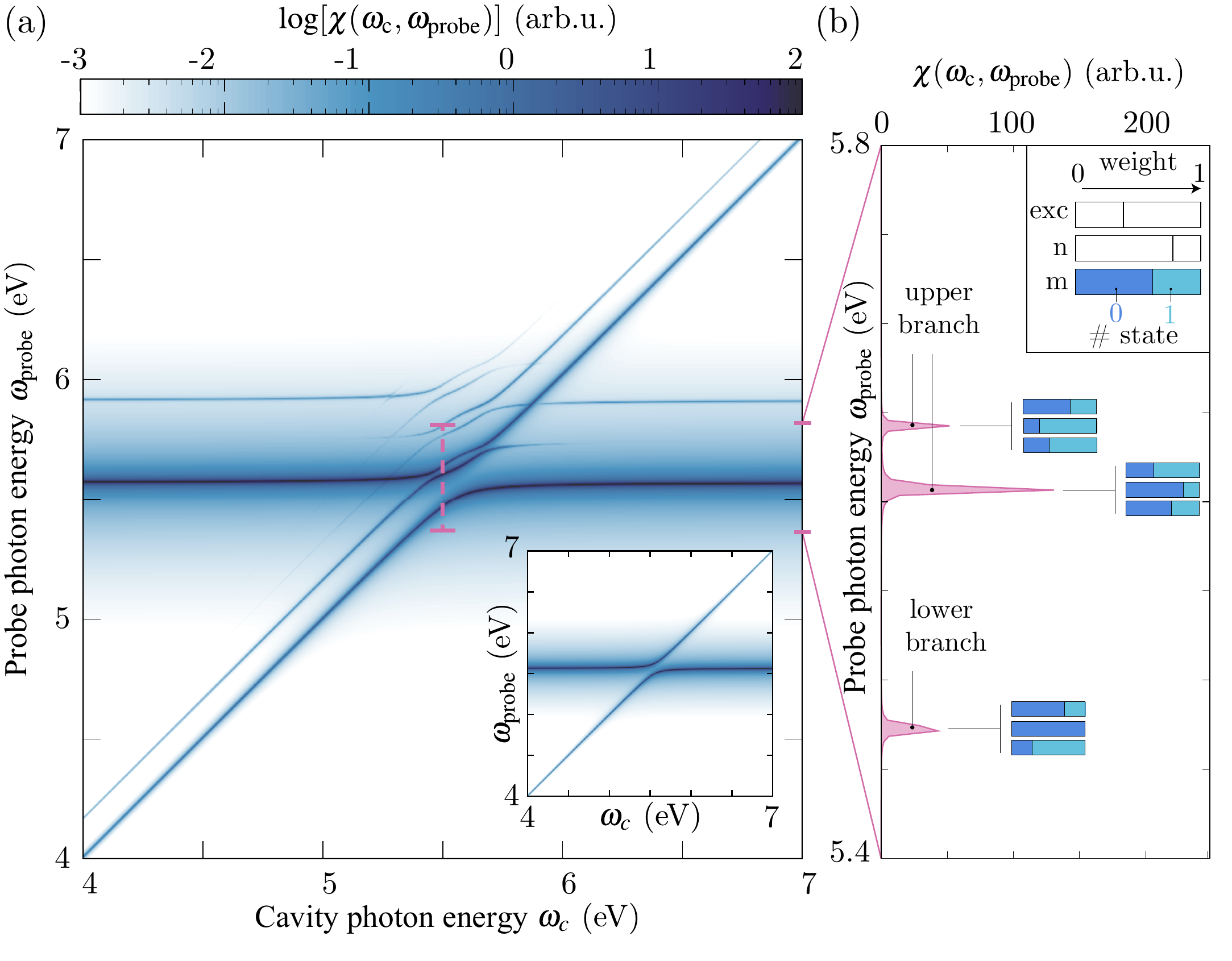}
   \caption{\label{fig:2}
   Computed optical spectrum of a monolayer hBN in a strongly coupled optical cavity as a function of cavity energy $\omega_c$ for the coupling strength $A_0=0.02$ a.u.. The optical response is shown in logarithmic scale, $\log[\chi]$. The characteristic avoided crossing of an exciton polariton is modulated by the additional hybridization of phonon modes in the material, giving composite states of exciton-photon-phonon character. The inset shows the exciton-polariton response in absence of phononic and phonoritonic coupling. The dashed line indicates a cut through the spectrum that is analysed in panel (b): each of these peaks are transitions from the ground state to an excited state of the phonoriton Hamiltonian (see text).  }
\end{figure}

{\it Results.} In order to unravel the spectral features of the phonoriton, we calculate the matter component of the full optical response $\chi(\omega_c,\omega_{\rm probe})$, a quantity which can be experimentally measured. Within linear response theory, the matter response is given by \cite{DavisArXiv2018,ruggenthaler2018quantum}: $\chi(\omega_c,\omega_{\rm probe}) = \sum_{I}\frac{\mathcal{M}^{\rm p}_{0I}\mathcal{M}^{\rm p}_{I0}}{\omega_{\rm probe}-E^{\rm p}_{I}(\omega_c)+E^{\rm p}_{0}(\omega_c)+i\eta}
$, where $\mathcal{M}^{\rm p}_{IJ}=\langle\Psi^{\rm p}_I|\hat{p}|\Psi^{\rm p}_{J}\rangle$ and $\eta$ is an artificial broadening which in the following is kept fixed but can in principle be tuned to reproduce thermal broadening of the spectral features. 

Fig.~\ref{fig:2} reports the ab-initio optical response of hBN embedded in a cavity with tunable frequency $\omega_c$ and coupling strength $A_0=0.02$~a.u.. Here, one can observe the appearance of a strong optical signal between the lower and upper exciton-polariton branches, as compared to the bare exciton-polariton case (c.f. inset Fig.~\ref{fig:2}(a)). The additional spectral features show the presence of the phononic series that interfere with the upper exciton-polariton branch yielding a strong signal modulation that we identify as phonoritonic resonances.

The formation of phonoritons opens up new optical transition channels for the phonon sidebands that are otherwise dipole forbidden in the uncoupled states, because all excitations couple to the same cavity photon. The cavity photon acts as a glue introducing further coupling between excitons and phonon modes. 
We note that the direct exciton-phonon coupling adds spectral features associated with phonon-replicas, at energies above the upper exciton-polariton branch, which are not as intense as the phonoriton signal and indeed can only be observed in logarithmic scale.

Analysing the ab-initio matrix elements in the Hamiltonian, Eq.~\ref{eq:QED_corrected}, we can single out the LO mode as the one responsible for the formation of phonoritons in hBN, because it is the only phonon which has a non-zero phonoriton matrix element, $\mathcal{Z}_{\rm LO} =-0.02+0.014i$~a.u.. For the exciton-phonon coupling both LO and TO modes have similar values ($\mathcal{G}_{\rm LO}=0.004+0.002i$~a.u.,  $\mathcal{G}_{\rm TO}=0.002+0.004i$~a.u.) while the ZO mode has a negligible value. Since the TO mode does not contribute to the phonoriton and is degenerate with the LO mode, it only contributes to the intensity of the exciton-phonon spectral signal and is therefore omitted in the calculations. The exciton-photon coupling is instead $\mathcal{M}_{01}=0.06+0.10i$~a.u..

To disentangle which mixing channels are enabled by the cavity we have dissected the contribution of the different particles, exciton, phonons and photons, to the phonoritonic features that are discernible in a linear scale, for a cavity frequency of $\omega_c=5.5$~eV, c.f. Fig.~\ref{fig:2}(b). To quantify the degree of hybridization we define a particle weight by taking the square of the phonoritonic state $\Psi^{\rm p}_I$ responsible of a given spectral feature and trace out all the degrees of freedom, except for one kind of particle. For example the excitonic weight is computed as $w_I^{\rm{exc}} = \sum_{ n m} |\Psi^{\rm p}_{I, inm}|^2$, and similarly for the LO phonon mode and for the cavity photon, shown in Fig.~\ref{fig:2}(b). 
States in different polaritonic branches are characterized by different excitonic, phononic and photonic content: the lower branch is mainly made of the ground electronic state and one photon, while the upper branch, which is split in two, contains excitonic states as well as both phononic and photonic number states up to $n,m=1$. This peak splitting reveals the formation of a phonoriton. In particular, the peak at $\omega_{\rm probe} \approx 5.61$~eV has a major contribution of the excitonic state and the $n,m=0$ phononic and photonic states while the peak at $\omega_{\rm probe} \approx 5.64$~eV is mainly made by the ground electronic state and the $n,m=1$ phononic and photonic states. 

\begin{center}
\begin{figure}
   \includegraphics[width=\columnwidth]{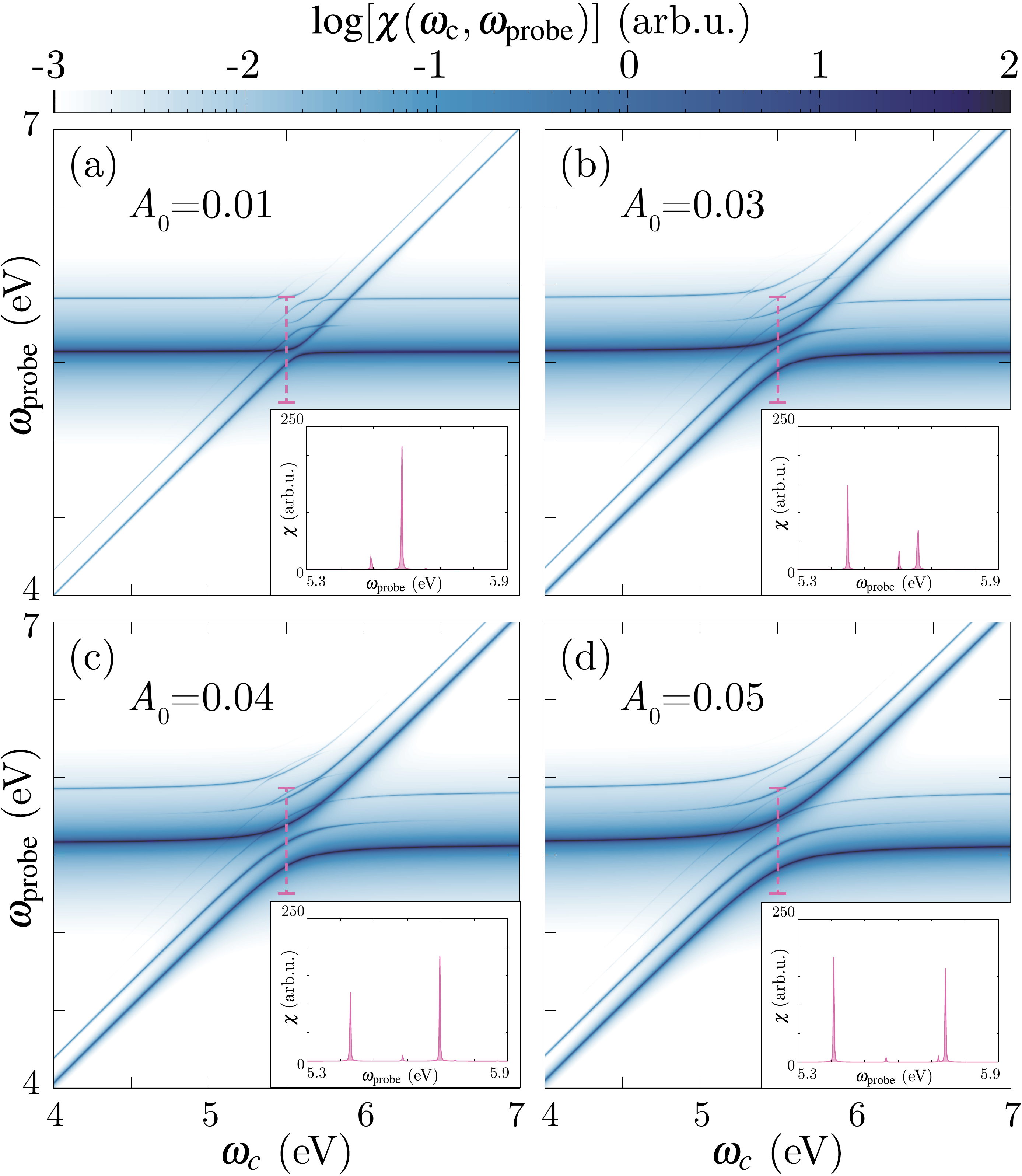}
   \caption{\label{fig:3}
    Computed optical spectrum of a monolayer hBN in a strongly coupled optical cavity as a function of cavity energy $\omega_c$ for different coupling strength $A_0$ in atomic units and in logarithmic scale. Dashed lines indicate cuts that correspond to the spectra shown in linear scale in the insets.
    } 
\end{figure}
\end{center}

The fact that the upper polaritonic branch participates in the phonoriton formation hints at the great potential of tunability of the phonoritonic response. Indeed by changing the cavity frequency, the Rabi-splitting can be tuned to a different phonon number or in more complex materials to other phonon modes to selectively involve them in the phonoriton formation. We stress that the coupling strength is within experimental reach as demonstrated in Ref.~\cite{FlattenSciRep2016} and we have shown in Ref.~\cite{Latini:2019bz} that the cavity parameters can be tuned, for example, by changing the size of the device and/or altering the dielectric environment. 
Finally, to highlight the role of cavity-matter coupling and hence the Rabi splitting in the realization of the phonoritonic states we calculated the optical response of hBN for different coupling strengths, c.f. Fig.~\ref{fig:3}. Somewhat counter-intuitively, a stronger coupling does not necessarily enhance the phonoriton, but instead the comparison of the different couplings in Fig.~\ref{fig:3} and Fig.~\ref{fig:2}(a) show that for a clear phonoriton signal the phonon line needs to be close to resonance with the Rabi splitting and since a low phonon number states have a stronger phonoritonic coupling, a large Rabi splitting can wash out the phonoriton features. 

The phonoriton features described here rely on a phonoritonic coupling term that is not negligible when compared to the exciton-photon coupling. Due to mass dependence of the phonoritonic coupling (see definition) it is expected, as shown for hBN, that crystals composed of lighter atoms are more likely to host phonoritonic quasiparticles. This is confirmed by calculations for the transition metal dichalcogenide WS$_2$, which has heavier elements than hBN, where we have found the phonoriton features to be much less pronounced.

A relevant experimental challenge for the realization of our proposal is the fabrication of a cavity whose photon energy is high enough to be in resonance with the exciton in hBN. Indeed a Fabry-Perot type of cavity, with a fundamental mode of $5.5$~eV, would require a distance between the mirrors of $~225$~nm. It is however feasible to employ higher cavity modes of existing cavity designs in the IR range~\cite{Liu:2015iy}. This approach would be the equivalent to using higher harmonics of a laser. The theory reported above directly extends to the case of higher modes as long as the $A_0$ is properly scaled as $A_0\propto 1/\sqrt{n}$ where $n$ is the mode number, under the assumption that the cavity modes are independent, as shown in Ref.~\cite{Latini:2019bz}. 
Another aspect to consider is the choice of the encapsulating dielectric material, such that it does not interact with the cavity (UV) photons, possible candidates are oxides like HfO$_2$. 
An alternative pathway towards the realization of phonoritonic states could be to change the host material to one with a smaller bandgap that has a small phonon effective mass and retains strong excitonic features in the visible range: graphane~\cite{Elias:2009jd} and graphene oxide are promising candidates. Using two different layers as active material is yet another option, where one  provides the excitons and the other one provides the phonons: a possible candidate with such features is the WSe$_2$-hBN bilayer which shows a strong coupling of the WSe$_2$ exciton  with the ZO phonons of hBN~\cite{Jin:2017dj}. Finally, while the presented phonoritonic description is explicited for the case of zero momentum particles, it can be directly extended to finite momentum and be for example used to investigate the potential of highly tunable surface acoustic waves (SAW) for the formation of phonoritons~\cite{cho2005bragg, cerda2010polariton, cerda2013exciton}.

{\it Conclusion.} We have proposed a general first-principle framework to account for lattice motion effects in the QED Hamiltonian and we rediscovered a three-way quasiparticle, the phonoriton. We have predicted the effect of this exciton-phonon-photon quasiparticle on the optical response of monolayer hBN in an optical cavity. The standard exciton-polariton branch picture is enriched by bright features originating from hybridized phonon states which are glued together by the photon of the cavity. We have also shown that via the cavity-matter coupling strength one can potentially control the ingredients of the phonoriton by involving different phonons or other modes, condensate phonoritons or realize chiral phonoritons in chiral cavities\cite{Hubener:2020fm}.

\begin{acknowledgments}
We thank Emre Ergecen for constructive discussions. We acknowledge financial support from the European Research Council(ERC-2015-AdG-694097). Grupos Consolidados (IT1249-19) and the Cluster of Excellence 'CUI: Advanced Imaging of Matter' of the Deutsche Forschungsgemeinschaft (DFG) - EXC 2056 - project ID 390715994. The Flatiron Institute is a division of the Simons Foundation. S. L. acknowledges support from the Alexander von Humboldt foundation. Work at MIT was supported by the US Department of Energy, BES DMSE and by the Gordon and Betty Moore Foundation’s EPiQS Initiative grant GBMF9459.
\end{acknowledgments}

\end{document}


\author{Simone~Latini}
\email{simone.latini@mpsd.mpg.de}
\affiliation{Max Planck Institute for the Structure and Dynamics of Matter and Center for Free Electron Laser Science, 22761 Hamburg, Germany}

\author{Umberto~De~Giovannini}
\affiliation{Max Planck Institute for the Structure and Dynamics of Matter and Center for Free Electron Laser Science, 22761 Hamburg, Germany}
\affiliation{Nano-Bio Spectroscopy Group,  Departamento de Fisica de Materiales, Universidad del País Vasco UPV/EHU- 20018 San Sebastián, Spain}

\author{Edbert J. Sie}
\affiliation{Department of Physics, Massachusetts Institute of Technology, Cambridge, MA 02139, USA}
\affiliation{Geballe Laboratory for Advanced Materials, Stanford University, Stanford, CA 94305, USA}

\author{Nuh Gedik}
\affiliation{Department of Physics, Massachusetts Institute of Technology, Cambridge, MA 02139, USA}

\author{Hannes~H\"ubener}
\affiliation{Max Planck Institute for the Structure and Dynamics of Matter and Center for Free Electron Laser Science, 22761 Hamburg, Germany}

\author{Angel~Rubio}
\email{angel.rubio@mpsd.mpg.de}
\affiliation{Max Planck Institute for the Structure and Dynamics of Matter and Center for Free Electron Laser Science, 22761 Hamburg, Germany}
\affiliation{Nano-Bio Spectroscopy Group,  Departamento de Fisica de Materiales, Universidad del País Vasco UPV/EHU- 20018 San Sebastián, Spain}
\affiliation{Center for Computational Quantum Physics (CCQ), The Flatiron Institute, 162 Fifth avenue, New York NY 10010.}

\title{Supporting Information:
Phonoritons in a monolayer hBN optical cavity}
\date{\today}

\maketitle

\section{1. Derivation of the QED Phonoritonic Hamiltonian}

To describe the dressing of excitons by phonons and cavity photons from first principles, we consider the fundamental many-body Hamiltonian in (non-relativistic) quantum electrodynamics (QED). We consider here a single photonic cavity mode corresponding to the lowest momentum mode perpendicular to the cavity mirrors and in-plane electric field polarization. The generalization to multi-modes is not expected to change the conclusions as shown in Ref.~\cite{Latini:2019bz}. Within the velocity gauge, the QED problem can be written as follows:
\begin{equation}
\label{eq:QED}
\hat{H}_{\rm QED} = \hat{H}_{\rm{el}} + \omega \hat{b}^\dagger \hat{b} +  \frac{A_0^2}{2}(\hat{b}^\dagger - \hat{b})^2 + A_0 \sum_{rs\bf{k}}\left(\langle\phi_{r\bf{k}} | \hat{e}\cdot\hat{p} | \phi_{s\bf{k}}\rangle \hat{c}^\dagger_{r\bf{k}}\hat{c}_{s \bf{k}} \hat{b}^\dagger +  h.c.\right),
\end{equation}
where $\hat{b}^\dagger$ and $\hat{b}$ are the photon creation and annihilation operators respectively, $\hat{H}_{\rm{el}}$ is the many-body electronic Hamiltonian, $\hat{c}^\dagger_{r\bf{k}},\hat{c}_{s \bf{k}}$ are the electronic creation and annihilation operators in the $\{\phi_r\}$ basis (with $r$ band indices and $\bf{k}$ wavevectors in the first Brillouin zone), $\hat{p}$ the single particle momentum operator, $\hat{e}$ the photon field polarization and $A_0=\sqrt{\frac{1}{2\pi c~ a}}$, with $a$ the area of the unit cell, is the amplitude of the vector potential.
In order to derive the coupling of this Hamiltonian to lattice degrees of freedom we consider the parametric dependence of the Hamiltonian on the phonon displacements of the atoms $\{{\bf R}_{{\bf q}\alpha}\}$, corresponding to the phonon mode $\alpha$ with momentum ${\bf q}$. To first order in this displacement, the Hamiltonian can be expanded as
\begin{equation}
    \hat{H}_{\rm QED} \simeq \hat{H}_{\rm QED, ~eq} + \left.\frac{\partial \hat{H}_{\rm QED}}{\partial {\bf R}_{{\bf q}\alpha}}\right|_{\rm eq}{\bf R}_{{\bf q}\alpha}
\end{equation}
where the subscript "eq" indicates the ${\bf R}_{{\bf q}\alpha}=0$ condition which corresponds to no phonon displacements. The first term in the equation is the usual polaritonic Hamiltonian, while the second term describes its coupling to the lattice. Since the coupling of photons to the electronic structure occurs via the creation/annihilation of neutral excitations (electron-hole pairs), one can approximate the many-body electronic eigenstates of the Hamiltonian by the excitonic ones and therefore diagonalize the electronic component, $\hat{H}_{\rm{el}}\left|\Psi_{i}^{\rm{exc}}\right\rangle\simeq\epsilon_i^{\rm{exc}}\left|\Psi_{i}^{\rm{exc}}\right\rangle$~\cite{Latini:2019bz}. This gives matrix elements like
\begin{equation}
\label{eq:QEDexpansion}
\left\langle\Psi_{i}^{\rm{exc}}\right|\hat{H}_{\rm QED}\left|\Psi_{j}^{\rm{exc}}\right\rangle = \left\langle\Psi_{i, {\rm eq}}^{\rm{exc}}\right|\hat{H}_{\rm QED, ~eq}\left|\Psi_{j, {\rm eq}}^{\rm{exc}}\right\rangle 
+ \left\langle\Psi_{i, {\rm eq}}^{\rm{exc}}\right|\left.\frac{\partial \hat{H}_{\rm QED}}{\partial {\bf R}_{{\bf q}\alpha}}\right|_{\rm eq}\left|\Psi_{j, {\rm eq}}^{\rm{exc}}\right\rangle {\bf R}_{{\bf q}\alpha},
\end{equation}
Since in this work we focus on the effect of the $\Gamma$-phonons ($q=0$), in the following we drop the sum over momenta. Nevertheless the results can be generalized to finite $q$, which is necessary when dealing, for example, with exciton coupling with surface acoustic waves or any other finite momentum excitation. 
The first term of Eq.~\ref{eq:QEDexpansion} can be expressed as follows:
\begin{equation}
\left\langle\Psi_{i, {\rm eq}}^{\rm{exc}}\right|\hat{H}_{\rm QED, ~eq}\left|\Psi_{j, {\rm eq}}^{\rm{exc}}\right\rangle = \left[\epsilon_{j, {\rm eq}}^{\rm{exc}} +\omega \hat{b}^\dagger \hat{b} + \sum_{\alpha}\Omega_{\alpha} \hat{a}^\dagger_{\alpha} \hat{a}_{\alpha} + \frac{A_0^2}{2}(\hat{b}^\dagger + \hat{b})^2\right]\delta_{ij} +
A_0 \left(\mathcal{M}^{\rm exc}_{ij} \hat{b}^\dagger + \mathcal{M}^{\rm exc *}_{ji} \hat{b}\right)
\end{equation}
where $\mathcal{M}^{\rm exc}_{ij} = \langle\Psi_{i, {\rm eq}}^{\rm{exc}}|\sum_{rs\bf{k}} \langle\phi_{r\bf{k}, {\rm eq}} | \hat{e}\cdot\hat{p} | \phi_{s\bf{k}, {\rm eq}}\rangle \hat{c}^\dagger_{r\bf{k}}\hat{c}_{s \bf{k}}|\Psi_{j, {\rm eq}}^{\rm{exc}}\rangle $ are excitonic matrix elements of the bilinear electron-photon coupling, the same as the ones introduced in Ref.~\cite{Latini:2019bz}. For completeness we also introduced the energy term for the phonons with $a_\alpha^\dagger$ and $a_\alpha$ being the creation and annihilating operators for the mode with index $\alpha$.

\begin{figure}
  \centering
   \includegraphics[width=0.7\columnwidth]{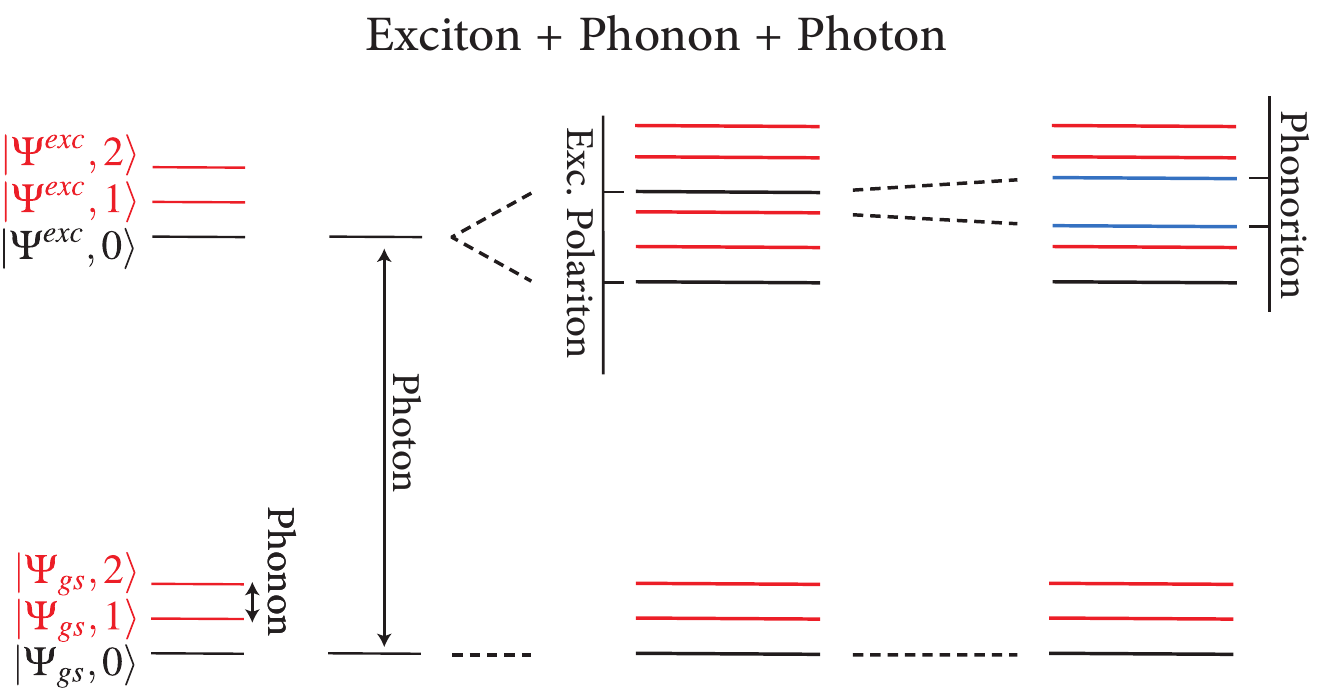}
   \caption{\label{fig:S1}
   {\bf Three-particle-hybridization in a Phonoriton.}  Energy level representation of the hybridization of exciton, phonon and photons involves mutual dressing of the modes: the cavity photon (left panel), coupling directly to the exciton, creates exciton-polariton states (black line, middle panel), which through the phonoritonic coupling are further mixed with phonons to yield the phonoritonic states (blue line, right panel).
   }
\end{figure}

The second term in Eq.~\ref{eq:QEDexpansion} gives origin to two terms, the standard exciton-phonon coupling and the phonoritonic coupling introduced in the main text. Indeed we can rewrite this term as:
\begin{equation}
\label{eq:excphonandphnrt}
\begin{split}
\left\langle\Psi_{i, {\rm eq}}^{\rm{exc}}\right|\left.\frac{\partial \hat{H}_{\rm QED}}{\partial {\bf R}_{\alpha}}\right|_{\rm eq}\left|\Psi_{j, {\rm eq}}^{\rm{exc}}\right\rangle {\bf R}_{\alpha}=& \left\langle\Psi_{i, {\rm eq}}^{\rm{exc}}\right|\left.\frac{\partial \hat{V}}{\partial {\bf R}_{\alpha}}\right|_{\rm eq}\left|\Psi_{j, {\rm eq}}^{\rm{exc}}\right\rangle {\bf R}_{\alpha} + \\
&+A_0 \left\langle\Psi_{i, {\rm eq}}^{\rm{exc}}\right|\left.\frac{\partial }{\partial {\bf R}_{\alpha}}\left[\sum_{rs\bf{k}}\left(\langle\phi_{r\bf{k}} | \hat{e}\cdot\hat{p} | \phi_{s\bf{k}}\rangle \hat{c}^\dagger_{r\bf{k}}\hat{c}_{s \bf{k}} \hat{b}^\dagger +  h.c.\right)\right]\right|_{\rm eq}\left|\Psi_{j, {\rm eq}}^{\rm{exc}}\right\rangle {\bf R}_{\alpha}
\end{split}
\end{equation}
where $\hat{V}$ is the electrostatic potential generated by the nuclei. 
We identify the first term on the right hand side as the exciton-phonon coupling:
\begin{equation}\label{eq:EPC}
\mathcal{G}^{\rm exc}_{ij, \alpha} \equiv \left\langle\Psi_{i, {\rm eq}}^{\rm{exc}}\right|\left.\frac{\partial \hat{V}}{\partial {\bf R}_{\alpha}}\right|_{\rm eq}\left|\Psi_{j, {\rm eq}}^{\rm{exc}}\right\rangle =  \langle\Psi_{i, {\rm eq}}^{\rm{exc}}|\sum_{rs\bf{k}} g_{rs\bf{k}}^{\alpha}\hat{c}^\dagger_{r\bf{k}}\hat{c}_{s \bf{k}}|\Psi_{j, {\rm eq}}^{\rm{exc}}\rangle,
\end{equation}
with $g_{rs\bf{k}}$ the standard single-particle electron-phonon matrix elements~\cite{Giustino.2017}. The second term on the right hand side of Eq.~\ref{eq:excphonandphnrt} contains instead the phonoritonic matrix elements which essentially arise from the variation of the momentum matrix element with respect to the lattice vibration:
\begin{equation}
\mathcal{Z}_{ij, \alpha}^{\rm exc}=\sqrt{\frac{1}{2M_{\alpha}\Omega_{\alpha}}} \langle\Psi_{i, {\rm eq}}^{\rm{exc}}|\sum_{rs\bf{k}} \left.\frac{\partial}{\partial{\bf R}_{\alpha}}\langle\phi_{r\bf{k}} | \hat{e}\cdot\hat{p} | \phi_{s\bf{k}}\rangle\hat{c}^\dagger_{r\bf{k}}\hat{c}_{s \bf{k}}\right|_{\rm eq}|\Psi_{j, {\rm eq}}^{\rm{exc}}\rangle,
\end{equation}
where we have included the prefactor coming from the canonical transformation of the phononic displacement into creation and annihilation operators, i.e. ${\bf R}_{\alpha}= \sqrt{\frac{1}{2M_{\alpha}\Omega_{\alpha}}} (\hat{a}^\dagger_{\alpha} + \hat{a}_{\alpha})$.
Finally we can rewrite the full QED Hamiltonian 
coupled linearly to lattice degrees of freedom as
\begin{equation}\label{eq:QED_LR}
\begin{split}
\langle\Psi_{i}^{\rm{exc}}|\hat{H}|\Psi_{M}^{\rm{exc}}\rangle =
&\left[\epsilon_{j}^{\rm{exc}} +\omega \hat{b}^\dagger \hat{b} + \sum_{\alpha}\Omega_{\alpha} \hat{a}^\dagger_{\alpha} \hat{a}_{\alpha} + \frac{A_0^2}{2}(\hat{b}^\dagger + \hat{b})^2\right]\delta_{ij} + \sum_{\alpha}\left(\mathcal{G}^{\rm exc}_{ij, \alpha} \hat{a}^\dagger_{\alpha} + \mathcal{G}^{\rm exc *}_{ji, \alpha} \hat{a}_{\alpha}\right) \\
&+ A_0 \left(\mathcal{M}^{\rm exc}_{ij} \hat{b}^\dagger + \mathcal{M}^{\rm exc *}_{ji} \hat{b}\right) + A_0 \sum_{\alpha}\left(\mathcal{Z}^{\rm exc}_{ij, \alpha} \hat{b}^\dagger + \mathcal{Z}^{\rm exc *}_{ji, \alpha} \hat{b}\right)\left(\hat{a}^\dagger_{\alpha} + \hat{a}_{\alpha})\right.
\end{split}
\end{equation}
This is the central equation of the main text and we refer to the eigenstates of such Hamiltonian as the phonoriton quasiparticle states. As sketched in Fig.~\ref{fig:S1}, the photon acts by creating replicas of the purely electronic and phononic states resulting in an optical Stark effect or Rabi splitting where the originally uncoupled states hybridize to yield phonoriton states that have simultaneously electronic, phononic and photonic character. \\

In this work we consider hBN and in particular the non-dispersive (momentum independent) excitonic states localized around the K-points of the Brillouin zone. These excitonic states occur in a non-hydrogenic series, which is accurately described by a two-band BSE, where only a single valence and a single conduction band are taken into account \cite{LatiniPhysRevB2015}. With this simplification the matrix elements appearing in Eq.~(\ref{eq:QED_LR}) have one of the two following structures:
\begin{align}\label{eq:ME_GS}
\mathcal{M}^{\rm exc}_{0j} &=\sum_{\rm{\bf{k}}} A_{\rm{\bf{k}}}^{j}\langle \phi_{v\bf{k}} | \hat{e}\cdot\hat{p} | \phi_{c\bf{k}}\rangle,\\
\mathcal{M}^{\rm exc}_{ij} &= \sum_{\rm{\bf{k}}} A_{\rm{\bf{k}}}^{i*}A_{\rm{\bf{k}}}^{j}\left[\langle\phi_{c\bf{k}} | \hat{e}\cdot\hat{p} | \phi_{c\bf{k}}\rangle - \langle \phi_{v\bf{k}} | \hat{e}\cdot\hat{p} | \phi_{v\bf{k}}\rangle\right]\label{eq:ME_EXC}.
\end{align}
In the equations above we have expressed the excitonic states as a linear combination of singly excited electronic determinants, where the coefficients of the linear combination are given by the solution of the BSE \cite{RohlfingPhysRevLett1998,RohlfingPhysRevB2000,OnidaRevModPhys2002},
$
|\Psi_{j}^{\rm{exc}}\rangle = \sum_{cv\bf{k}} A_{cv\bf{k}}^{j} \hat{c}^\dagger_{c\bf{k}}\hat{c}_{v \bf{k}}|\Psi_0\rangle
$
with $A_{cv\bf{k}}^{n}$ the BSE coefficients, or envelope functions and $c$ and $v$ indices running over conduction and valence bands respectively. The electronic groundstate $|\Psi_0\rangle \equiv |\Psi_{\rm{j}=0}^{\rm{exc}}\rangle$ can instead be written as a single determinant of only valence states.
Similar expressions as in Eq.~\ref{eq:ME_GS} and ~\ref{eq:ME_EXC} can be readily obtained for the exciton-phonon matrix elements:
\begin{align}\label{eq:ME_GS_ph}
\mathcal{G}^{\rm exc}_{0j} &=\sum_{\rm{\bf{k}}} A_{\rm{\bf{k}}}^{j} g_{vc\bf{k}},\\
\mathcal{G}^{\rm exc}_{ij} &= \sum_{\rm{\bf{k}}} A_{\rm{\bf{k}}}^{i*}A_{\rm{\bf{k}}}^{j}\left[g_{cc\bf{k}} - g_{vv\bf{k}}\right]\label{eq:ME_EXC_ph}.
\end{align}

\section{2. Electron-Phonon Coupling and Excitonic Problem} 
The ab-initio quantities entering the Hamiltonian in Eq.~2 solved in the main paper require two main calculations: electron-phonon matrix elements and exciton energies and wavefunctions.

The electron-phonon matrix elements have been calculated from using the octopus code~\cite{TancogneDejean:2020ek}, by displacing the lattice according to the phonon eigenmodes and subsequently projecting the variation of the electronic potential into the unperturbed Kohn-Sham basis. 

The excitonic energies and wavefunctions are instead obtained by solving the Bethe-Salpeter Equation (BSE~\cite{OnidaRevModPhys2002}) on a LDA basis using the GPAW code \cite{MortensenPhysRevB2005,EnkovaaraJPhysCondMat2010}. The BSE is an equation that can be derived within many-body perturbation theory and takes into account the many-body screened interaction between the electron and the hole forming the excitons. In practice, such an equation can be cast in a two-particle Hamiltonian represented in the space of valence and conduction bands. Given the strongly localized nature of the 1s exciton in hBN (the one considered in the manuscript), it is enough to include the last valence and the first conduction band~\cite{wirtz2008comment, blase1995quasiparticle, galambosi2011anisotropic}. The screened interaction is then calculated at the RPA level by including all the bands up to $~100 eV$. The BSE is then solved by exact diagonalization on a $60\times60$ k-points grid and the excitonic energies and the corresponding wavefunctions $A^{i}_{\bf k}$ are obtained. In order to properly describe the electronic gap of $7.12$~eV as reported in Ref.~\cite{haastrup2Dmaterials2018}, we have corrected the LDA bandgap by applying a scissor operator.